\documentclass[fleqn,10pt,twocolumn]{wlscirep}

\providecommand{\abs}[1]{\lvert#1\rvert}
\newcommand{\onlinecite}[1]{\nocite{#1}\citenum{#1}}

\newcommand*{\ket}[1]{\left\lvert #1 \right\rangle}
\newcommand*{\bket}[2]{\left\langle \, #1 \,|\, #2 \, \right\rangle}

\title{A 3D investigation of delocalised oxygen two-level defects in Josephson junctions}

\author[1,*]{Timothy C. DuBois}
\author[1]{Salvy P. Russo}
\author[1]{Jared H. Cole}
\affil[1]{Chemical and Quantum Physics, School of Applied Sciences, RMIT University, Melbourne, 3001, Australia}

\affil[*]{tim.dubois@tcqp.science}

\begin{abstract}
Environmental two-level systems (TLS) have been identified as significant decoherence sources in Josephson junction (JJ) based circuits. For such quantum devices to be functional, the removal or control of the TLS is a necessity. Understanding the microscopic origins of the ‘strongly coupled’ TLS type is one current path of investigation to that end. The delocalized oxygen model suggests the atomic position of an oxygen atom is spatially delocalized in the oxide forming the JJ barrier. In this report we extend this model from its previous 2+1D construction to a complete 3D description using a Wick-rotated time-dependent Schr\"{o}dinger equation to solve for time-independent solutions in three dimensions. We compute experimentally observable parameters for phase qubits and compare the results to the 2+1D framework. We devise a Voronoi classification scheme to investigate oxygen atoms delocalizing within strained and non-strained crystalline lattices, as well as realistic atomic positions a JJ amorphous tunnel barrier constructed in previous density functional studies.
\end{abstract}
\begin{document}

\flushbottom
\maketitle

\thispagestyle{empty}

\section*{Introduction}

Decoherence sources such as the environmental two-level system (TLS)~\cite{Dutta1981, Shnirman2005} are presently a major hurdle to the realisation of operational superconducting qubits and other Josephson junction based quantum devices.
Many properties of the TLS have been probed experimentally, demonstrating relatively long decoherence times whilst showing them to be stable and controllable~\cite{Simmonds2004, Neeley2008, Shalibo2010, Lupascu2009, Lisenfeld2010, Grabovskij2012, Lisenfeld2015}.
However, identifying their true microscopic nature remains an open problem. Controllable superconducting qubit architectures (charge, flux and phase) make it possible to study single `strongly coupled' TLSs~\cite{Neeley2008, Lupascu2009, Lisenfeld2010}, as opposed to weakly coupled ensembles of TLSs that may be responsible for $1/f$ noise~\cite{Dutta1981,Schriefl2006}.
Although the Standard Tunneling Model (STM) \cite{Anderson1972, Phillips1972} can be used to explain all existing experimental results, to improve devices we need a detailed microscopic understanding of these defects.

Many microscopic models have now been proposed: delocalised oxygens~\cite{DuBois2013,DuBois2015}, surface state interactions~\cite{Choi2009,Lee2014}, polaron dressed electrons~\cite{Agarwal2013} and alien species (primarily hydrogen)~\cite{Jameson2011, Holder2013, Gordon2014}; all of which are possible decoherence sources and/or explain the origin of strongly coupled TLSs.
Better fabrication methods, higher vacuums and more robust qubit designs have historically suppressed or diminished the response of such noise sources~\cite{Vion2002, Martinis2005, Koch2007, Schreier2008, Houck2008}.
However, continuing down the path of a perfectly clean but amorphous dielectric may no longer be the optimal direction. Whilst crystalline layers are difficult to construct in this environment, they are possible and show a dramatic decrease (of up to 80\%) in visible TLSs over their amorphous counterparts~\cite{Oh2006}.

Such a decrease could be explained via our previous work on the delocalised oxygen model~\cite{DuBois2013,DuBois2015}, where we suggest the origin of some TLS defects are oxygen atoms in a spatially delocalised state due to the non-crystalline nature of the dielectric layer.
As such, this is a concrete microscopic ensemble of the classic atomic tunneling model: typically cited as motivation for the STM (which is more general, as the tunneling object need not be a single atom).

Our previous model discussed a simplified configuration space using low dimensional arguments, culminating in a 2+1D framework where a cage of six aluminium atoms surrounded a central oxygen atom in three dimensions and calculated observations based on the oxygen delocalising on a two dimensional plane.
The dimensional simplicity of this model allowed a direct matrix diagonalisation approach to solve a time independent Schr\"{o}dinger equation, which due to memory limits could not be extended into a complete 3D representation of a TLS defect.

Here we introduce an alternative method for this model which can obtain a complete three dimensional picture of the delocalised oxygen ground and excited states.
We use a Wick-rotated time-dependent Schr\"{o}dinger equation to solve for time-independent solutions in three dimensions (see Methods for more detail).
As the work in this report is a higher dimensional extension of the 2+1D model, its terminology has been retained herein.
It is suggested the reader be familiar with the discussions in Ref \onlinecite{DuBois2015}, although a short summary of the nomenclature is outlined below.
Using this more powerful technique, we are able to move beyond a simplistic cubic representation and investigate true lattice configurations and realistic amorphous positions from our DFT studies~\cite{DuBois2013,DuBois2015a}.

Two defect `types' are investigated, both of which stem from deforming the crystalline lattice of corundum: the low temperature and pressure phase of $\mathrm{Al_2O_3}$ (the  closest Al--O bond distance in this phase is $\sim\!1.85$ \AA).
Defect type A; where the aluminium--oxygen bond distance is shortened, forcing the oxygen to occupy one of two off-axis positions, and defect type B; where the opposite occurs: the aluminium--oxygen bond distance is lengthened, allowing two preferred oxygen positions, on-axis.

The trigonal lattice structure of this crystalline material and the amorphous nature of a JJ barrier give rise to many complications for a microscopic model, due to potential contributions from many neighbouring atoms to a possible defect site.
Therefore, to fully understand the response to these interactions, we start with a very simplistic cluster of atoms and add complexity once an understanding of rudimentary behaviour is achieved.

Defining the oxygen position to be at an origin, two introduced aluminium atoms can be considered as pairs ($x = -X, \: +X$ = \{$\pm 1.85$ \AA\}) lying on a cardinal axes; which are identified as $\abs{X}$ or simply referred to as the `defect pair'.
Displacing $\abs{X}$ equidistantly from this origin (i.e.\ moving away from optimal crystalline configuration) will yield either an A or B type defect, depending on the direction of displacement.
$\abs{Y}$ and $\abs{Z}$ describe aluminum pairs in the $y$ and $z$ directions respectively.

Eigenenergies of a system are frequently discussed in a relative fashion using a convention where $E_{ij} = E_j-E_i$, such that the ground state ($E_0$) to first excited state ($E_1$) energy splitting is defined as $E_{01}$.

\section*{Results}

\begin{figure}[ht]
  \centering
  \includegraphics[width=\linewidth]{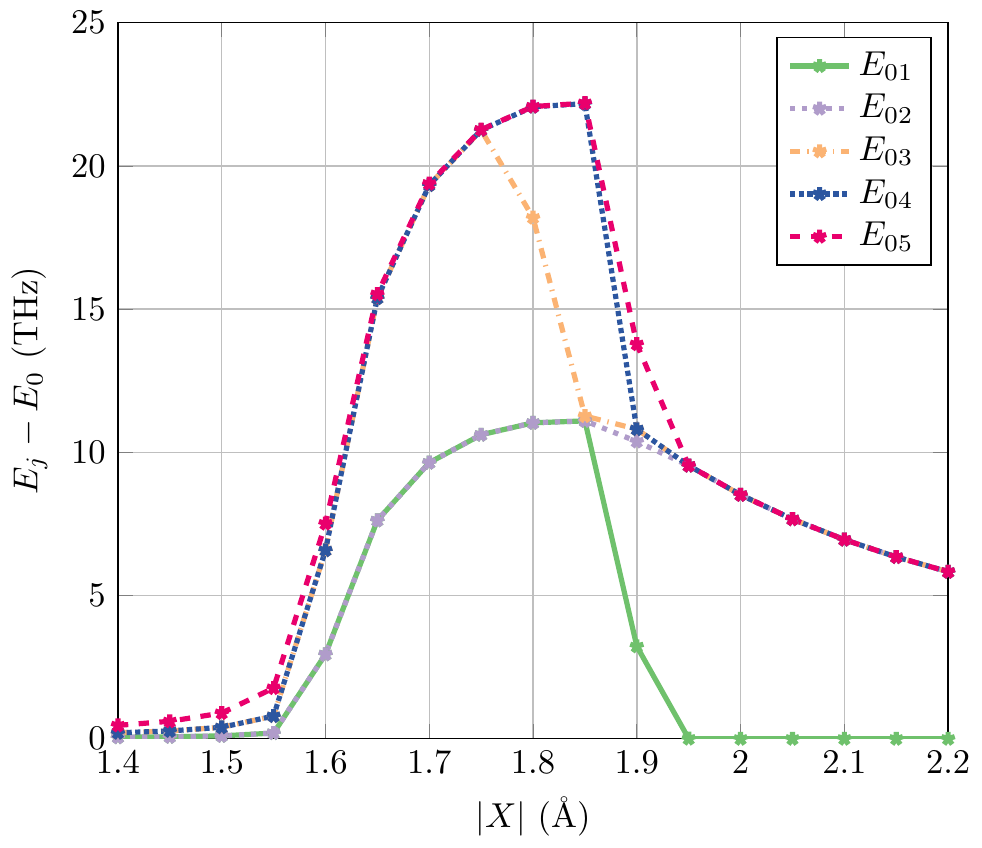}
  \caption{Eigenspectrum of a three atom system Al--O--Al, over a varying distance separation $\abs{X}$. Each excited state has been normalised with the ground state, which shows a clear $E_{01}$ degeneracy at large separation distance - indicative of a B type defect. An intermediate region exists as the separation distance decreases, which is approximately centered about the optimal Al--O bond distance of corundum ($1.85$ \AA). At small separation distances, the eigenspectrum is distributed over comparatively small energy levels, but is not degenerate (see Figure \ref{fig:lift}).}
  \label{fig:spect3d}
\end{figure}

The capability to investigate oxygen delocalisation in 3D, using realistic atomic positions within an amorphous layer of $\rm{Al_2O_3}$ generates an extensive state space.
Hence we start the investigation using a simple three atom system comprising of an oxygen atom and a confining aluminium pair: $\abs{X}$.
An eigenspectrum of the ground and five lowest excited energy levels of this system is generated by increasing or decreasing the pair separation distance from the corrundum lattice distance $1.85$ \AA\, which is depicted in Figure \ref{fig:spect3d} (for comparison with the 2D model, see Figure 2 in Ref \onlinecite{DuBois2015}).

In this figure an (an)harmonic division about $\abs{X} \sim 1.85$ \AA\ separates two regions with dissimilar properties.
In the region where $\abs{X} > 1.95$ \AA, a degenerate $E_{01}$ is observed and as the separation distance has increased, can be identified as a possible B type defect.
The second region ($\abs{X} < 1.55$ \AA) has a tightly bound eigenspectrum compared to the rest of the map although this yields no degeneracy.

\begin{figure*}[ht]
  \centering
  \includegraphics[width=\linewidth]{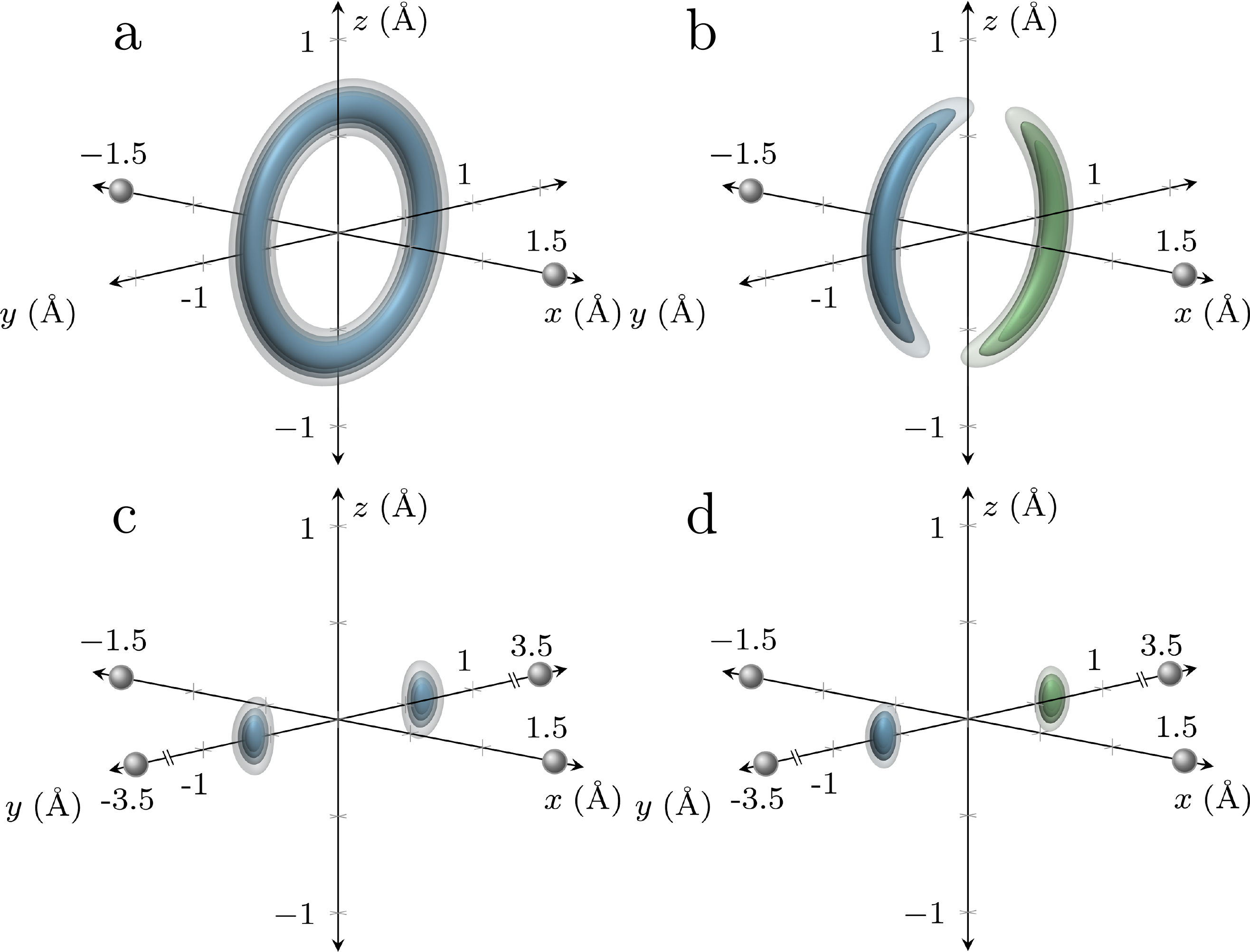}
  \caption{Ground (left) and first excited (right) state wavefunctions of an Al--O--Al chain with $\abs{X} = 1.5$ \AA\ (aluminium atoms are depicted as gray spheres). Top axes (a \& b) show a unique ground state, the $E_{12}$ pair is degenerate in this case. Two more aluminium atoms are introduced at $\abs{Y} = 3.5$ \AA\ on the bottom axes (c \& d), which causes the degeneracy to lift to the $E_{01}$ pair.}
  \label{fig:lift}
\end{figure*}

In three dimensions, the potential minima manifests as a sphere around the location of an atom. As a consequence the three atom Al--O--Al chain produces a unique, rotationally symmetric ground state which is reminiscent of an oxygen interstitial defect in crystalline germanium~\cite{Artacho1995} and depicted in Figures \ref{fig:lift}a\&b.
Comparatively, this region in the 2D case indicated the presence of an A type defect, as the 2D potential minima ring would be projected onto a plane, collapsing a degree of freedom.

Two other observations concerning the 2D to 3D transition is the dramatically reduced total energy values as wavefunctions are no longer artificially confined, and the higher excited states in the B type region now exhibit a quad degeneracy rather than two split doubly degenerate pairs - again because states are not artificially confined in the extra dimension.

This simplistic three atom description can now be built upon to understand the interactions of a delocalised oxygen in a three dimensional space.
A confining pair is introduced in an additional dimension to the configuration in Figures \ref{fig:lift}a\&b, and the unique ground state is lifted to a degenerate $E_{01}$ pair; observing an A type TLS defect.
Figures \ref{fig:lift}c\&d illustrate this effect with a confining pair $\abs{Y} = 3.5$ \AA.
This distance is an arbitrary choice, as the shift occurs even at the cut-off limits imposed by the potential portion of the model~\cite{Streitz1994} (see Methods).
In other words, any additional potential contribution which does not share the axial symmetry of the Al--O--Al chain will lift the degeneracy and result in a TLS.

\subsection*{Oxygenic Orbitals}

A complete picture of the ground and excited state wavefunctions in 3D is now possible, which allows us to further investigate the properties of the two-level system and illustrate the importance of crystalline dielectrics in future Josephson junction devices.

\begin{figure}[ht]
  \centering
  \includegraphics[width=\linewidth]{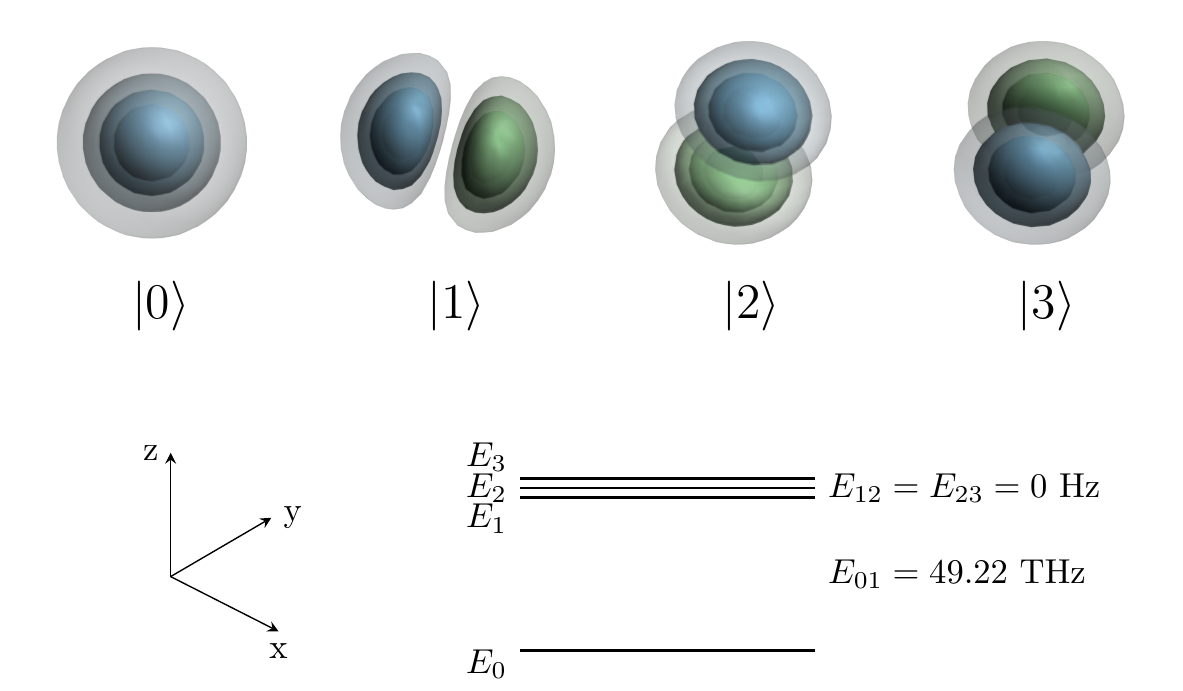}
  \caption{Wavefunctions of an oxygen confined by six equidistant aluminium atoms at $\abs{X}\!=\!\abs{Y}\!=\!\abs{Z} = 1.6$ \AA. This configuration exhibits a unique ground state (as shown in the energy level diagram) and hence is considered to be spatially localised.}
  \label{fig:lineharm}
\end{figure}

A confined harmonic state can exist with many atomic configurations, and as with the low dimensional model, setting variables symmetrically is one of these cases.
Figure \ref{fig:lineharm} depicts the four lowest energy wavefunctions of an oxygen with six confining aluminium atoms: $\abs{X}\!=\!\abs{Y}\!=\!\abs{Z} = 1.6$ \AA.
Here the oxygen atom has no additional local minima to occupy, identifying this configuration as spatially localised, the harmonic approximation holds and the system is not considered as a TLS candidate.

\begin{figure}[ht]
  \centering
  \includegraphics[width=\linewidth]{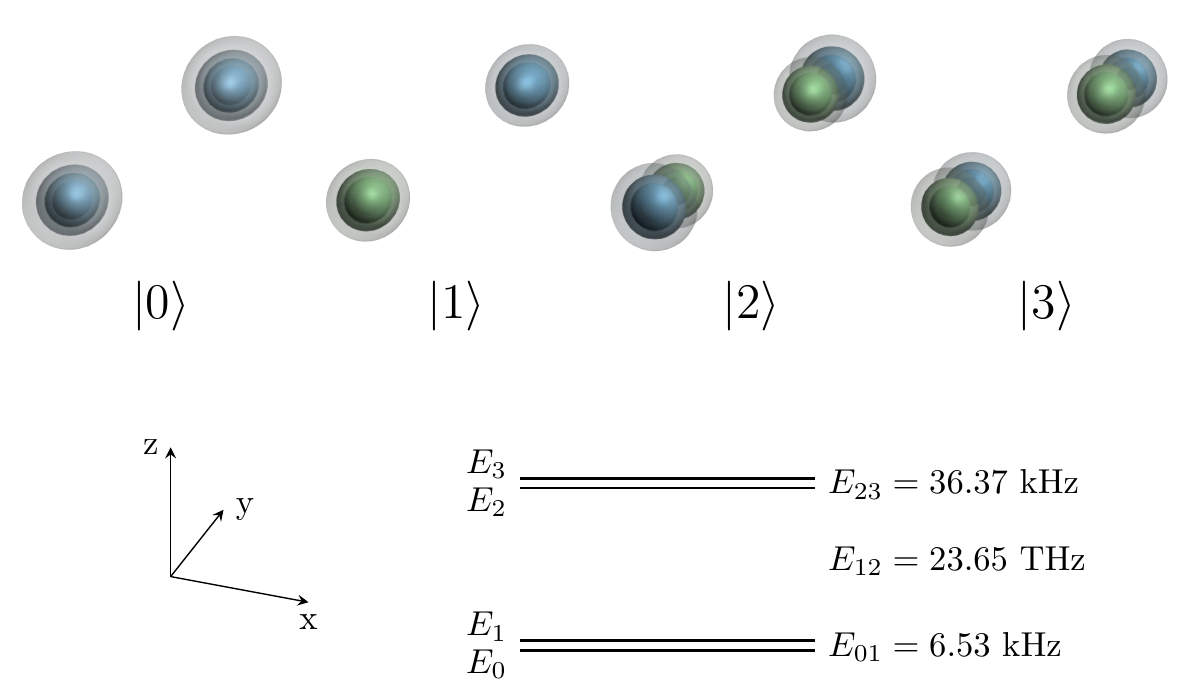}
  \caption{Wavefunctions of an oxygen confined by six aluminium atoms, where the symmetry of Figure \ref{fig:lineharm} has been broken. $\abs{X}\!=\!\abs{Z} = 1.6, \abs{Y} = 2.1$ \AA, which manifests as a B type defect.}
  \label{fig:lineb}
\end{figure}

From this localised case, we extend the $\abs{Y}$ confinement out to $2.1$ \AA.
Using the Figure \ref{fig:spect3d} eigenspectrum we can predict this configuration to be a B type defect; where the $\abs{Y}$ separation distance is increased, and an oxygen dipole forms parallel to the $y$ axis.
This is indeed the case as shown in Figure \ref{fig:lineb}.

\begin{figure}[ht]
  \centering
  \includegraphics[width=\linewidth]{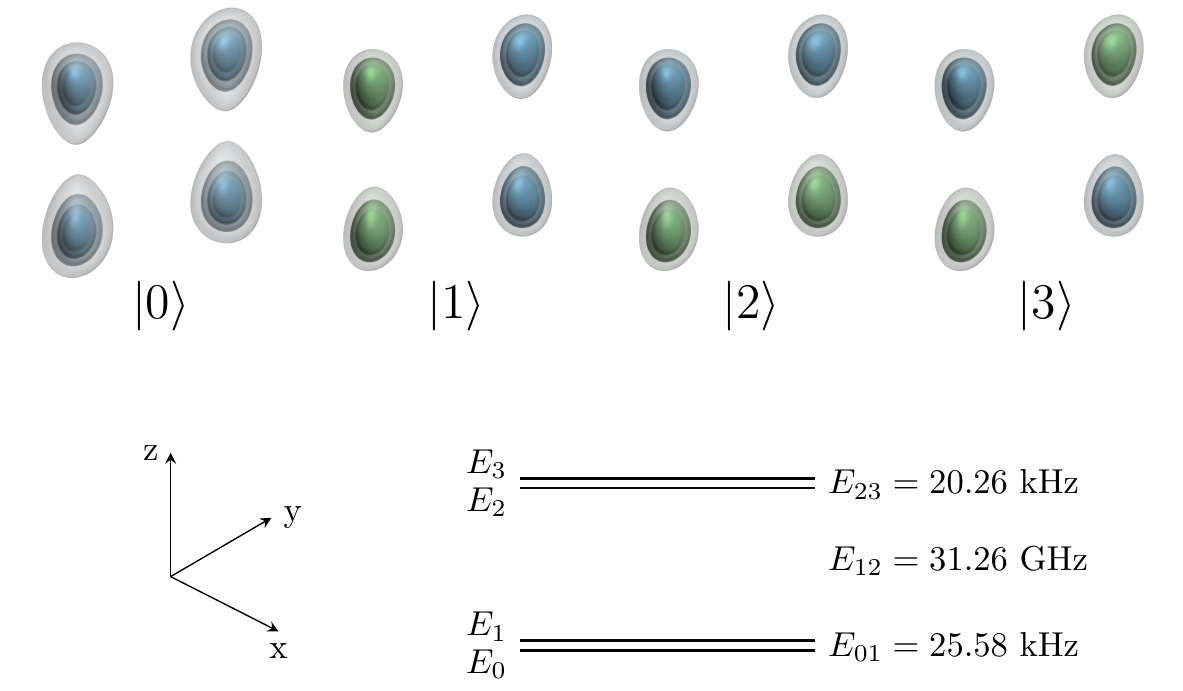}
  \caption{Wavefunctions of an oxygen confined by six aluminium atoms, where the confinement configuration is studiously pathological. $\abs{X} = 1.6, \abs{Y} = 2.1, \abs{Z} = 2.0$ \AA, yielding a quad degeneracy in the ground state due to the $E_{12}$ splitting existing below the superconducting gap ($\sim\! 100$ GHz, see text).  }
  \label{fig:linequad}
\end{figure}

A particularly complex phenomenon emerged from the low dimensional model: quad degenerate ground states, where the energy difference between two degenerate pairs $E_{01}$ and $E_{23}$ approach the difference of the pairs themselves, i.e.\ $E_{01} = E_{23} \approx E_{12}$.
Figure \ref{fig:linequad} shows a configuration expressing this behaviour.

The discussion in Ref \onlinecite{DuBois2015} indicated that this form of degeneracy has a low probability of occurrence compared to A or B type defects.
Additionally, if $E_{12} \geq 100$ GHz the higher states ($E_{2}$ and $E_{3}$) can be ignored completely as energies of this magnitude dissociate Cooper-pairs, hence can be viewed as an operational upper bound for Josephson junction devices.
Splitting energies in 3D are much smaller than the 2+1D case, which suggests that the probability of these defects being experimentally visible may be higher than predicted with the low dimensional model based solely on this measure.
However, a comprehensive analysis of the three dimensional configuration space would need to be undertaken before real estimates of this behaviour can be given.

Another interpretation of this phenomena has to do with the energy resolution of the coherent
single-pulse resonant driving spectroscopy~\cite{Lisenfeld2010} used to obtain qubit-TLS coupling ($S_{max}$) values.
The minimum resolvable energy using this process is approximately $1$ MHz, which may indicate that $S_{max}$ anti-crossing measurements are manifestations of qubit-TLS couplings at the $E_{12}$ level, if a configuration such as Figure \ref{fig:linequad} is scrutinised. $E_{01}$ and $E_{23}$ splitting levels would be hidden in their degenerate subspace and manifest as an effective two-level defect with $\widetilde{E}_{01} = 31.26$ GHz.

\subsection*{A type defects and dipole considerations}

Whilst adding additional confinement pairs causes the unique, rotationally symmetric ground state of $\abs{X} < 1.55$ \AA\ (Figures \ref{fig:lift}a\&b) to become degenerate; the third dimension yields a complication in the dipole measurement for the A type region.

\begin{figure}[ht]
  \centering
  \includegraphics[width=\linewidth]{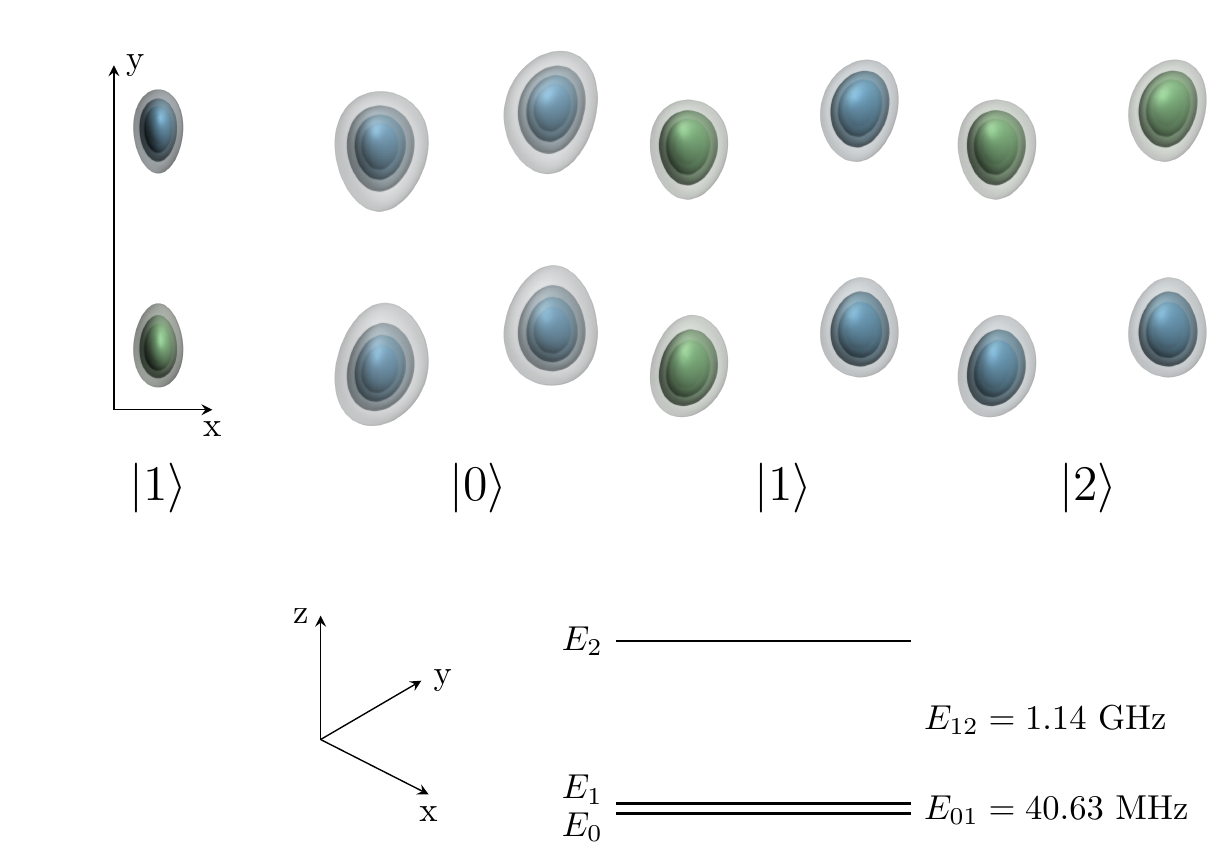}
  \caption{Wavefunctions of an A type defect with confinement $\abs{X} = 1.53, \abs{Y} = 1.97, \abs{Z} = 1.95$ \AA\ which yields dominant dipole in the $y$ direction with a strength of $0.26 \; e$\AA. Leftmost axis shows an $xy$ projection of the first excited state to compare with 2+1D results in Ref \onlinecite{DuBois2015}.}
  \label{fig:atypey}
\end{figure}

Consider a system with parameters $\abs{X} = 1.53, \abs{Y} = 1.97, \abs{Z} = 1.95$ \AA, illustrated in Figure \ref{fig:atypey}.
This system exhibits TLS behaviour, with $E_{01} = 40.63$ MHz, and a dipole strength in $y$ of $0.26 \; e$\AA: perpendicular to the confining $x$ axis as expected.
The leftmost axis of Figure \ref{fig:atypey} shows a 2D projection of the first excited state, illustrating no major differences in the response of the 3D and 2+1D models.

\begin{figure}[ht]
  \centering
  \includegraphics[width=\linewidth]{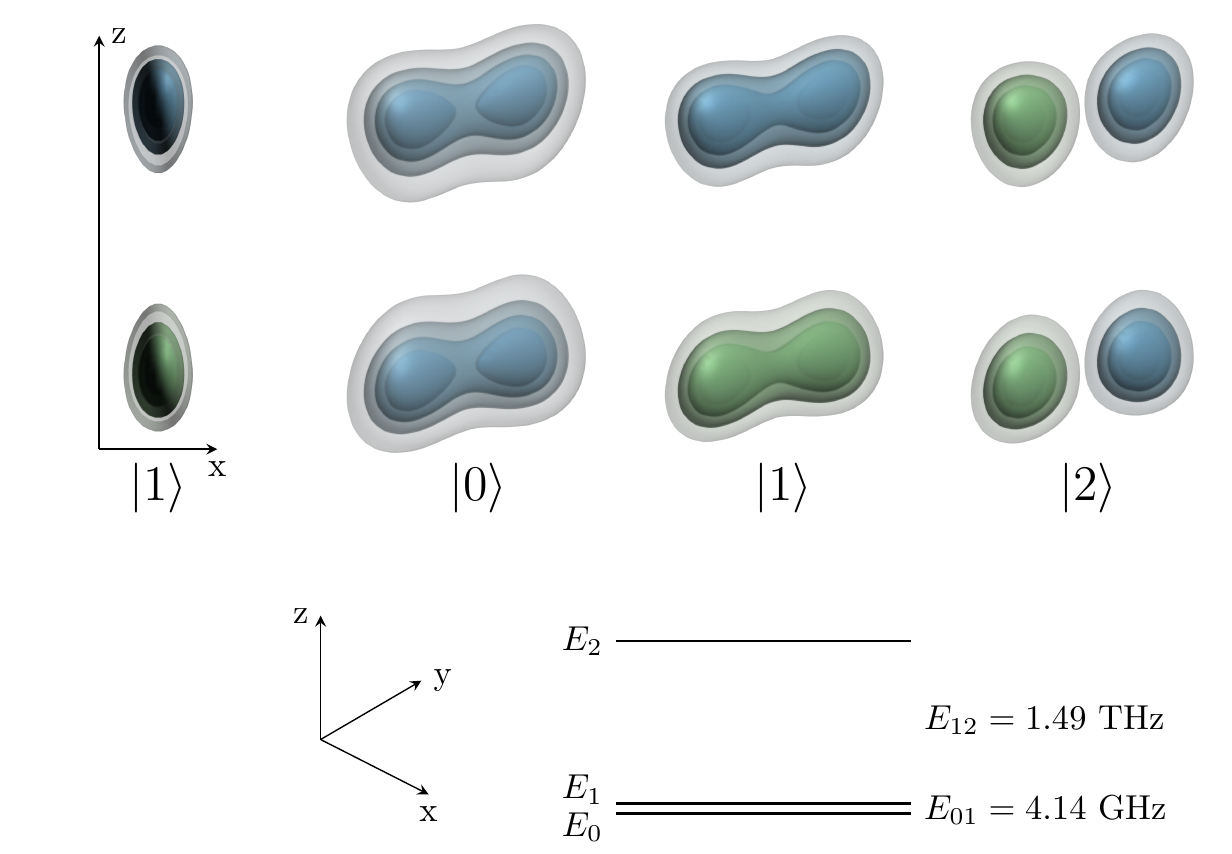}
  \caption{Wavefunctions of an A type defect with confinement $\abs{X} = 1.53, \abs{Y} = 1.9, \abs{Z} = 1.95$ \AA\ which yields dominant dipole in the $z$ direction with a strength of $0.21 \; e$\AA. Leftmost axis shows an $xz$ projection of the first excited state to compare with 2+1D results in Ref \onlinecite{DuBois2015}.}
  \label{fig:atypez}
\end{figure}

However, a small change in the $y$ confinement alters the system in a non-trivial manner.
Moving $\abs{Y}$ from $1.97$ to $1.9$ \AA\ (for example) crosses a bifurcation in state space.
As $\abs{Z}$ is now the least confining pair, the dominant dipole direction flips to $z$ as shown in Figure \ref{fig:atypez}.

Ultimately this does not add any complexity to the model: minimal confinement in $y$ generates an A type defect with a dipole in $y$, perpendicular to $x$; minimal confinement in $z$ generates an A type defect with a dipole in $z$, perpendicular to $x$.
As strongly coupled defects are experimentally identified via avoided level crossings in qubit spectroscopic diagrams~\cite{Lisenfeld2010}, the dipole moment of the defect couples to the electric field across the junction~\cite{Martinis2005}.
Therefore, to be identified as a TLS, the defect must be aligned to this field.
It follows then, that the dipole directions for an A type defect in this model can be considered equivalent.
Additionally, as with B type and any other possible TLS defect, both A type alignments will not appear on a spectroscopic scan as an avoided level crossing if their dipole is perpendicular to the external field.

\subsection*{Defects in corundum}

Results such as Figures \ref{fig:spect3d} and \ref{fig:lineharm} identify configurations which exhibit harmonic eigenspectrums and localised oxygenic positions.
This behaviour is expected when an oxygen atom is confined appropriately and has no reason to tunnel.
One way to check this assumption (and consequently further validate the TLS model) is to calculate the potential landscape seen by a single oxygen atom from bulk corundum ($\alpha$--$\mathrm{Al_2O_3}$).
The TLS model, delocalising an oxygen across all potential space about its minimum energy position, should yield a localised, harmonic wavefunction positioned at the location where the oxygen was removed (\textit{i.e.}\ the lattice position associated with the vacancy).

An oxygen atom in a corundum lattice is identified by a Voronoi polytope (see Methods) and a nearest neighbour atom cluster is used to calculate the wavefunctions and energies of an oxygen atom located inside the potential landscape generated by the corundum lattice.
As predicted, the result is a localised oxygen atom locked into the lattice position of the vacancy, and the oxygenic orbitals behave in a similar fashion to the harmonic system calculated previously in Figure \ref{fig:lineharm} (data not shown).

\subsection*{Straining the Local Lattice}

In an attempt to put bounds on the amount of amorphousness required for a near-crystalline structure to exhibit TLS properties, we apply a localised strain tensor on the defect cluster and observe its effects.
Two strain ratios $\frac{\Delta L'}{L} = \mp 0.019$ \AA\ representing a compressive force along the $x$ axis, and a tensile force also along $x$ respectively, attempt to simulate a local strain from the localised Al--O separation distance of $1.85$ \AA\ in the direction of the A type and B type regions respectively (see Figure \ref{fig:spect3d}).

The original crystalline system has a ground to first excited state splitting of $E_{01} = 13.48$ THz.
Comparing the $E_{01}$ values of the strained results (compressive: $35.42$ THz, tensile: $34.84$ THz) to the crystal result, we see an increase in splitting energy (data not shown).
This response is opposite to the simple straining of Figure \ref{fig:spect3d} where the $E_{01}$ energy was a maximum at the crystalline Al--O separation distance.
This suggests that local Poissonian strain does not give rise to immediate TLS behaviour---at least along the sampled axis $x$.

One possible explanation of this result comes from our previous study concerning strain on the 2+1D model~\cite{DuBois2015}: there exists a preferred strain axis which responds rapidly in comparison to other directions, although this axis is difficult to acquire directly from a completely harmonic case.
It is therefore pertinent to investigate the amorphous configurations in more detail.

\subsection*{Defects in Josephson Junctions}

Applying the Voronoi classification scheme the myriad atomic configurations within the Josephson junction models constructed in Ref.~\onlinecite{DuBois2015a} allows us to investigate realistic amorphous clusters residing within a JJ tunnel barrier.

\begin{figure}[ht]
\centering
\includegraphics[width=\linewidth]{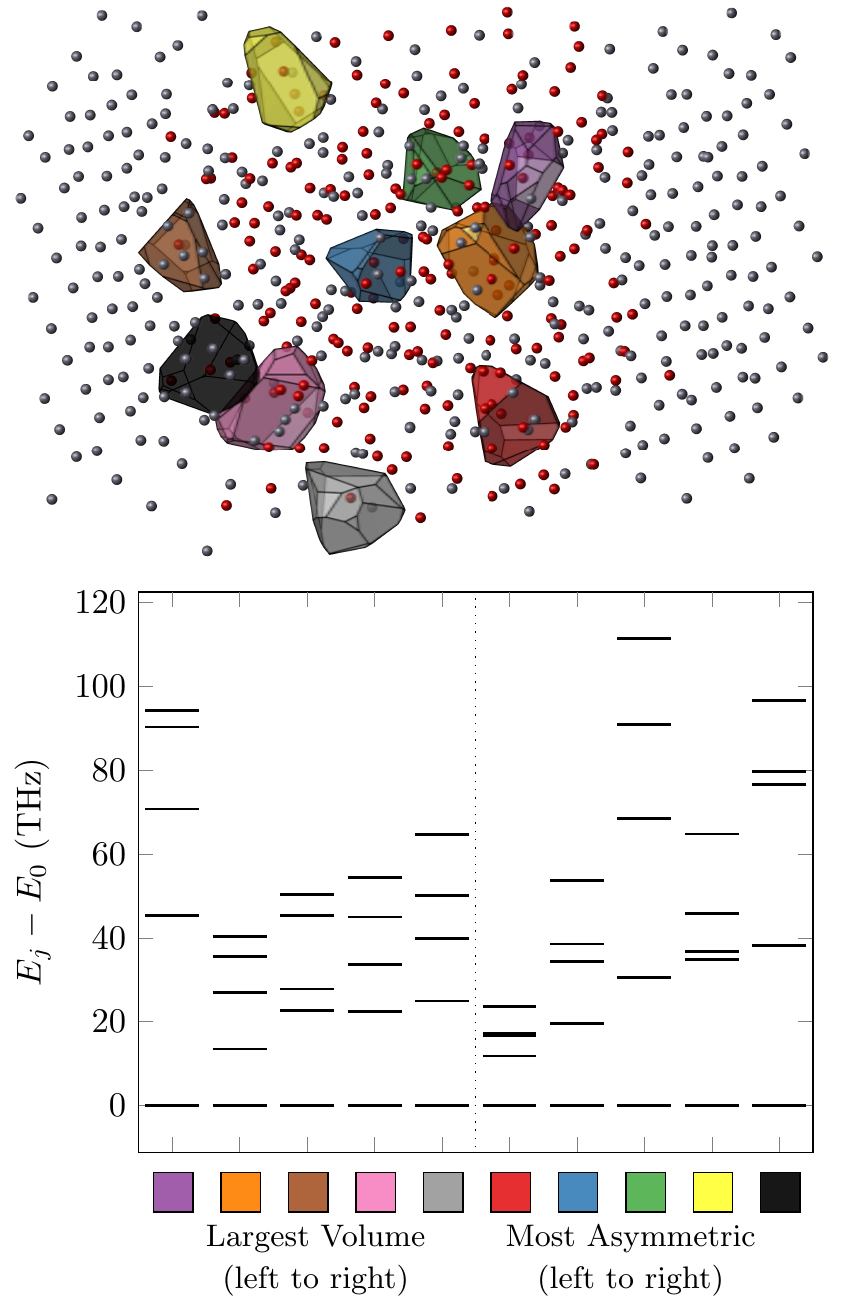}
\caption{Top: Oxygen atoms with the 5 largest Voronoi volumes and the 5 most asymmetric Voronoi cells are highlighted in within a Josephson junction comprised of aluminium (gray) and oxygen (red) atoms.  Bottom: Eigenspectrum of the ground state and four lowest excited state values for each of the candidate oxygen atoms.  }
\label{fig:tlsinjj}
\end{figure}

Using the structure with an $\mathrm{AlO_{1.25}}$ amorphous barrier with a density 0.75 that of corundum  we classify clusters for all oxygen atoms extant in the tunnel barrier.
It is posited that a large void space around an oxygen may allow it to delocalise over a large region and manifest as a B type defect (however, this is pure speculation).
The volume of each Voronoi cell can be calculated (as the cell is essentially a convex hull of some vertices describing a Delaunay tessellation) and we choose the five cells with the largest volume to investigate.
Another conjecture is that an asymmetric cell (obtained by some form of sheer in the local cluster configuration) yields TLS behaviour, thus a final cluster is investigated which has the most asymmetry -- identified using the length parameter of the spherocylinder formalism outlined in \onlinecite{Anikeenko2004}.
The five cells with the most asymmetry were also identified; these clusters and their properties are displayed in Figure \ref{fig:tlsinjj} along with the large volume candidates.

As with the cluster results of the crystalline and Poisson strain investigations, the few amorphous clusters studied failed to identify any TLS behaviour. Whilst each oxygen displays a different spectra, none of the $E_{01}$ splitting are degenerate and dipole strengths are no larger than $0.034 \; e$\AA\ (from the red cell, identified as the oxygen with the most asymmetry in its cluster) in any cardinal direction.

\section*{Discussion}

Using the direct diagonalisation method in three dimensions is seemingly intractable with current memory limitations, although this does not exclude investigations in the 3D domain.
The Wick-rotated time-dependent Schrödinger equation method has proved to be a valid tool to study this space, albeit with computational time disadvantages.

Whilst the 2+1D model allows us to investigate phase space in a much more flexible way, one of the major variations in the models is the final energy values, as the 3D model is no longer collapsed into fewer dimensions.
This result suggests that if one is attempting to identify exact values for a delocalised oxygen TLS, they should be utilizing the 3D method.
Contrarily, if trends are required, the 2+1D model provides the more efficient solution.
Whilst splitting energies and dipole results may not completely reflect the atomic positions of the surrounding clusters in this case, it is evident from discussions like those concerning Figures \ref{fig:atypey} and \ref{fig:atypez} that there are 2D to 3D equivalencies.

Using a Voronoi classification scheme to identify atom clusters which describes the local potential environment of an oxygen atom is shown to be a powerful method, describing a lattice position in crystalline corundum as well as the properties of any selected oxygen atom in a model junction.
The free parameter concern stated in Ref.~\onlinecite{DuBois2015} is somewhat mitigated through this process, as clusters with many atoms can be described with one Voronoi cell (clusters of up to 24 atoms are described in Figure \ref{fig:tlsinjj} for example).

The affect of local strain on a crystalline cluster observed behaviour which does not align with the simple strain investigations undertaken with the idealised 6 atom model, suggesting that the deformation axis chosen is not the preferred strain axis identified in Ref.~\onlinecite{DuBois2015} and that localised strain may not lead to TLS behaviour directly.

With such a small sample size it is difficult to draw any direct conclusions from the results in Figure \ref{fig:tlsinjj}.
Other classification metrics or methods of identification such as extending the Euclidian distance Voronoi classification discussed here to the Voronoi S-networks considered in \onlinecite{Anikeenko2004} may yield TLS behaviour with the present model.

However, the more probable explanation is that the surrounding atom clusters exert enough confinement energy to the oxygen atoms to keep them spatially localised, unlike the simpler 6 atom confinements.
One phenomena which may assist in successful TLS classification, is the inclusion of many body interactions (the C type defect in Figure 1 of Ref.~\onlinecite{DuBois2015}).
As oxygen atoms tunnel through potential barriers, the surrounding atomic neighbours may deform and create a multi-atom TLS candidate.
Such behaviour has been studied previously~\cite{Buchenau1984,Heuer1993,Daldoss1998,Trachenko2000,Reinisch2005}, and multi-body identification approaches have recently been investigated~\cite{Paz2014}.
Generalising the delocalised oxygen atom to a multi-body description is a topic for future work.

\section*{Methods}

To identify A and B defect types in either amorphous or strained crystalline systems, an effective single particle Hamiltonian can be used.
This simplifies the system by ignoring any time evolution properties or many-body interactions and yields the equation

\begin{equation}
    H = -\frac{\hbar^2}{2m_{oxy}}\nabla^2+V(\mathbf{r}),
    \label{eq:OHam}
\end{equation}

where $m_{oxy}$ is the mass of an oxygen atom and $V(\mathbf{r})$ is the potential due to the surrounding lattice.
This potential is modelled by the empirical Streitz-Mintmire potential~\cite{Streitz1994,Zhou2004}, which adequately captures the variable oxygen charge states occurring when oxygen is present in a predominantly metallic environment like a JJ metal-oxide interface.
A complete description of the parameters used can be found in Ref \onlinecite{DuBois2015} as the implementation remains consistent with this model.

To obtain full 3D eigenstates of this Hamiltonian we implement a Wick-rotated time-dependent Schr\"{o}dinger equation to imaginary time $t=i\tau$~\cite{Strickland2010},

\begin{align}
i \hbar \frac{\partial}{\partial t}\Psi(\mathbf{r},t) &= H\Psi(\mathbf{r},t)\\
\Rightarrow - \hbar \frac{\partial}{\partial \tau}\Psi(\mathbf{r},\tau) &= H\Psi(\mathbf{r},\tau)
\end{align}

which yields a general solution to the wavefunctions

\begin{equation}
\Psi(\mathbf{r},\tau) = \sum_{n=0}^\infty a_n\psi_n(\mathbf{r})e^{-E_n \tau}.\label{eq:psitau}
\end{equation}

Here, ${a_n}$ are coefficients based on initial conditions of the system where $n=0$ indicates the ground state, $n=1$ the first excited state \textit{etc}. and $E_n$ is the corresponding eigenenergy.
As $E_0 < E_1 < E_2 < \ldots$, evolving Eq.~\ref{eq:psitau} to large imaginary time will provide a good approximation to the ground state influenced by the time-independent potential $V(\mathbf{r})$.

This solution is obtained by numerically approximating the spatial derivatives with finite differences.
For stability and precision, we descritise the space using a $7$-point central difference method

\begin{multline}
f^{\prime\prime}(\mathbf{r}_0)\approx\\
\frac{2f_{-3}-27f_{-2}+270f_{-1}-490f_{0}+270f_{1}-27f_{2}+2f_{3}}{180h^{2}}
\end{multline}

where $f_k=f\left(\mathbf{r}_0+kh\right)$, calculated with a step size $h=0.01$ \AA. 

Excited states are obtained by Gram-Schmidt orthogonalisation~\cite{Gram1883, Schmidt1907}: choosing eigenfunctions within a degenerate subspace to be orthogonal to the system's basis.
For example, consider a converged, orthonormal ground state $\ket{\psi_0}$ and an initial, non-orthonormal guess for the first excited state $\ket{\psi^\prime_1}$.
Subtracting the projection of the excited state along the ground state from the initial guess yields a vector orthogonal to the ground state
\begin{equation}
\lvert\widetilde{\psi}^\prime_1\rangle = \ket{\psi^\prime_1}-\ket{\psi_0}\bket{\psi_0}{\psi^\prime_1}.
\end{equation}
This resultant vector can then be normalised and evolved through the same process as the ground state until convergence is achieved.
Similarly, the second excited state requires the same treatment, although projections along both $\ket{\psi_0}$ and $\ket{\psi_1}$ are needed.

An active repository of the software which implements these methods can be found via Ref \onlinecite{DuBois2015c}.

When dealing with the corundum crystal and amorphous JJ barrier clusters, we apply a Voronoi tessellation~\cite{Voronoi1908} to the lattice coordinates of each system using a Euclidean distance metric.
A polytope representing the convex hull encompassing an origin position symbolises the oxygen vacancy.
Atoms situated within polytopes sharing an edge with the origin polytope are selected as the model cluster.
These atoms constitute a first order screening of the surrounding lattice potential, representing an acceptable approximation to higher order screenings (\textit{e.g.} also including second nearest neighbours, which add appreciable computational intensity).
Clusters identified by this method use the Streitz Mintmire potentials generated by \texttt{GULP}~\cite{Gale2003}, which also account for monopole--monopole and self energy interactions to counteract any net charge generated by selecting an arbitrary cluster of atoms from a lattice with periodic boundaries.

Arbitrarily straining a corundum crystal, which has trigonal symmetry and six elastic constants~\cite{Bass1995}, will exert force on the structure in a non-trivial manner and act on many different crystal planes.
The materials science complications of this phenomena are well beyond the scope of this investigation, and instead we approximate a local strain by the Poisson effect~\cite{Poisson1829} using the values of the adiabatic bulk and shear moduli to obtain a Poisson's ratio $\nu = 0.234$.
This value is calculated as an equivalent isotropic polycrystalline aggregate to Sapphire (a mineral variety of corundum)~\cite{Gercek2007}.

The relationship
\begin{equation}
\left(1+\frac{\Delta L}{L}\right)^{-\nu} = 1-\frac{\Delta L'}{L}
\label{eq:strain}
\end{equation}
can be used to strain our atom cluster for example, with a length increase of $\Delta L$ in the $x$ direction, and a length decrease of $\Delta L'$ in the $y$ and $z$ directions.

\section*{Acknowledgements}
This research was supported under the Australian Research Council's Discovery Projects funding scheme (project number DP$140100375$). Computational resources were provided at the NCI National Facility systems at the Australian National University through the National Computational Merit Allocation Scheme supported by the Australian Government and the Victorian Partnership
for Advanced Computing (VPAC).

\section*{Author contributions statement}

T.D.\ Wrote the code base and initial draft, and generated figures. J.C.\ and S.R.\ provided direction and corrections. All authors analysed results and reviewed the manuscript.

%
%


\end{document}